# MODELO SEI$_1$I$_2$HRSVM APLICADO A LA PANDEMIA DE ENFERMEDAD POR CORONAVIRUS (COVID-19) EN PARAGUAY

**Juan A. González Cuevas***

*Facultad de Ingeniería, Universidad Nacional de Asunción, Campus San Lorenzo
**Correspondencia:** E-mail jgonzalez@ing.una.py

**Resumen:** En este artículo un modelo matemático es propuesto para investigar el actual brote del coronavirus (SARS-CoV-2) en Paraguay, describiendo las múltiples vías de transmisión y propagación en la dinámica de infección, haciendo seguimiento de los individuos susceptibles, expuestos, infecciosos, hospitalizados, recobrados, los que pierden su inmunidad, el rol del virus en el medio ambiente y los fallecidos por COVID-19 u otras razones. A fin de reflejar el impacto de las medidas de control en marcha adoptadas por el gobierno y la población, el modelo emplea tasas de transmisión variables que cambian con el estatus epidemiológico y condiciones ambientales. El modelo es validado y su aplicación es demostrada con datos disponibles públicamente.

**CONTENIDO**

# NOMENCLATURA

**VARIABLES DE ESTADO**

S(t) representa a los individuos susceptibles, es decir, aquellos que no han enfermado anteriormente y por lo tanto pueden resultar infectados al entrar en contacto con la enfermedad.

E(t) representa a los individuos expuestos, es decir, aquellos que portan la enfermedad pero que al hallarse en su periodo de incubación no muestran síntomas y no están en condición de infectar a otros, no requieren hospitalización, su aislamiento puede que no sea riguroso.

$I_1$(t) representa a los individuos inicialmente infectados no hospitalizados y por lo tanto en condiciones de transmitir la enfermedad. Los síntomas empeoran para una proporción de los primeros infectados, los cuales son hospitalizados, donde permanecen infecciosos. Los que presentan síntomas leves permanecen enfermos por la duración del segundo periodo infeccioso ($I_2$).

$I_2$(t) representa a los individuos del segundo periodo de infección y en condiciones de transmitir la enfermedad, presentan síntomas leves y se encuentran rigurosamente aislados, teniendo menor contacto con los expuestos o susceptibles, se recuperan pasando al grupo (R)

H(t) representa a los individuos hospitalizados; se asume que tienen 75% menor contacto con los individuos susceptibles. Del hospital pueden recuperarse (R) o morir (M).

R(t) representa a los individuos recobrados de la enfermedad, y que ya no están en condiciones de enfermar nuevamente ni de transmitir la enfermedad a otros por cierto tiempo, tras el cual pueden perder la inmunidad y volver a formar parte del grupo susceptible.

V(t) representa la concentración del virus SARS-CoV-2 en el medio ambiente.

Λ(t) representa al flujo de individuos que incrementa la cantidad de personas susceptibles. Como las medidas adoptadas por el gobierno incluyen cerrar las fronteras al ingreso de extranjeros, el incremental corresponde solo a los nacimientos sucedidos durante la pandemia o unos pocos connacionales retornando del extranjero.

M(t) representa al número de muertes naturales o inducidas por COVID-19

D(t) representa al virus desintegrado en el ambiente, por los que ya no contribuye a la propagación de la pandemia.

N(t) representa el número total de individuos en Paraguay

**PARAMETROS**

| Parámetro | Descripción | Valor promedio estimado | Dimensión | Fuente |
|---|---|---|---|---|
| $\Lambda$ | Tasa promedio de nacimientos e ingresos al país | 396.42 | $dia^{-1}$ | [1] |
| $\beta$ | Tasa de transmisión básica | 0.3 | $dias^{-1}$ | Ajuste con datos |
| $m_H$ | Reducción de transmisión en el hospital | 0.25 | Sin dimensión | [2] |
| $m_V$ | Factor de transmisión por virus en el medio ambiente | 0.45 | Sin dimensión | Ajuste con datos |
| c | Coeficiente de ajuste de transmisión | $5 \times 10^{-4}$ | Sin dimensión | Ajuste con datos |
| $1/\gamma$ | Tiempo promedio de infección inicial | 5.20 | días | [2] |
| $\mu$ | Tasa promedio de defunciones naturales | $1.49 \times 10^{-5}$ | $dias^{-1}$ | [1] |
| w | Tasa de progreso per cápita a través del estado de hospitalización | 1 / 4.96 | $dias^{-1}$ | [2] |
| f | Tasa promedio de pérdida de inmunidad | 1/1095 | $dias^{-1}$ | [3] |
| $1/\alpha$ | Tiempo promedio de incubación | 5.01 | días | [2] |
| $\lambda$ | Tasa de progreso per cápita a través del segundo estado de infección sin hospitalización | 1/10 | $dias^{-1}$ | [2] |
| $P_1$ | Proporción de población inicialmente infecciosa que se hospitaliza | 0.188 | Sin dimensión | [2] |
| $P_2$ | Proporción de población hospitalizada que mueren | 0.147 | Sin dimensión | [2] |
| $\varepsilon_1$ | Tasa de excreción del virus por personas inicialmente infecciosas | 1 | $personas^{-1} \times dias^{-1} \times ml$ | Ajuste con datos |
| $\varepsilon_2$ | Tasa de excreción del virus por personas infecciosas no hospitalizadas | 0 | $personas^{-1} \times dias^{-1} \times ml$ | [4] |
| $\varepsilon_3$ | Tasa de excreción del virus por personas hospitalizadas | 0 | $personas^{-1} \times dias^{-1} \times ml$ | [4] |
| $\sigma$ | Tasa de remoción del virus del ambiente | 1 | $dias^{-1}$ | [4][5] |

# I. Introducción

Desde la aparición de los primeros casos en Wuhan, China en diciembre de 2019, la infección del nuevo coronavirus (SARS-CoV-2) se ha propagado rápidamente por el mundo, ocurriendo los primeros casos en Paraguay a principios del mes de marzo de 2020, tras lo cual el gobierno ha implementado medidas de intervención y estrategias de mitigación.

Según lo reconocido por la Organización Mundial de la Salud [6], los modelos matemáticos desempeñan un papel clave en la información a los responsables políticos y de la salud para la toma de decisiones basadas en la evidencia. Estimaciones mediante simulaciones numéricas del potencial de transmisión, a menudo medido en términos del número de reproducción básico, el tiempo del pico del brote, y los valores y duración bajo medidas de intervención, proveen información crucial para determinar el potencial y la severidad de un brote [7].

En un modelo determinista la enfermedad puede infectar a los individuos aleatoriamente. Sin embargo, el número de infecciones se va haciendo cada vez más predecible conforme el tamaño de la población aumenta. Debido a esto los modelos deterministas son usados para tratar enfermedades que afectan a poblaciones grandes y a menudo surgen representados a través de ecuaciones diferenciales.

En 1927, W. O. Kermack y A. G. McKendrick crearon el modelo compartimental determinista SIR **[8]** que considera una enfermedad que se desarrolla a lo largo del tiempo y solo tres grupos de individuos, los susceptibles (S), los infectados (I) y los recobrados (R), donde cada grupo es mutuamente excluyente. Se han desarrollado varios modelos a partir del modelo SIR, como el modelo SEIS, el cual considera una nueva clase de individuos, los expuestos (E), suponiendo que un individuo que ha enfermado nunca obtiene inmunidad. El modelo SEIR es derivado del SEIS pero agregando a la población de recobrados. Recientemente, el modelo SEIRV [4] fue propuesto para modelar el brote de coronavirus en Wuhan, China, considerando el rol del virus esparcido en el medio ambiente (V) en la transmisión de la enfermedad y tasas de transmisión variables. Otro modelo propuesto considera a los individuos hospitalizados (H) [2]. En Paraguay previamente a este modelo Pastor [9] y Hyun [10] han utilizado un modelo SEIR para obtener predicciones de la pandemia del COVID-19, y Von Lucken ha estudiado el valor de reproducción efectivo del coronavirus en Paraguay [11].

En el actual trabajo, proponemos un modelo $SEI_1I_2HRSVM$ que combina los diversos aspectos de los modelos compartimentales mencionados anteriormente, haciendo seguimiento de los individuos susceptibles, expuestos, infecciosos, hospitalizados, recobrados que mantienen su inmunidad y los que la pierden pasado cierto tiempo, el rol del virus en el medio ambiente y los fallecidos (M) naturalmente, por COVID19 u otras razones. El modelo emplea tasas de transmisión variables para reflejar las medidas de control gubernamentales en Paraguay.

El resto del artículo está organizado como sigue: en la Sección II se presenta la metodología utilizada para modelar la pandemia del coronavirus, proveyendo una descripción de la estructura del modelo (II-a), las ecuaciones del modelo (II-b), el diagrama de transferencia (II-c), los coeficientes de transmisión variables (II-d) y el cálculo del número básico de reproducción (Ii-e); en la Sección III se proveen los resultados numéricos del trabajo, detallando el ajuste de curvas del modelo con datos reportados en Paraguay (III-a), la progresión de la pandemia con y sin adoptar medidas de control (III-b, III-c), el análisis de sensibilidad de parámetros y limitaciones del modelo (III-d); finalmente, las conclusiones a este trabajo son dadas en la Sección IV.

## II. Metodología
### a. Descripción de la estructura del modelo determinista

La población total consiste de siete clases: susceptibles (S), expuestos pero no infecciosos (E), primera clase de infecciosos ($I_1$), segunda clase de infecciosos ($I_2$), hospitalizados (H), recuperados (R), o muertos (M). Los individuos son considerados susceptibles hasta que entran en contacto con individuos infecciosos de ($I_1$), ($I_2$), (H) o el virus en el medio ambiente (V). Dado el contacto con un individuo infeccioso, la transmisión toma lugar con cierta probabilidad. Luego que la transmisión del virus ha ocurrido, los individuos susceptibles (S) pasan al grupo expuesto (E), donde permanecen un cierto periodo igual al tiempo promedio entre la infección y comienzo de infecciosidad (periodo de latencia, asumido igual al tiempo de incubación) Luego del periodo de latencia, los individuos pasan al primer periodo infeccioso ($I_1$). Los síntomas empeoran para algunas personas de $I_1$, quienes son hospitalizados (H), donde permanecen infecciosos. Los individuos que no son hospitalizados permanecen con síntomas leves por la duración del segundo periodo infeccioso ($I_2$). De ($I_2$), los individuos se recuperan (R) tras una progresión menos severa de la enfermedad. Del hospital, los individuos se recuperan o mueren (M). Asumimos que los individuos en el hospital tienen 75% menos contacto con individuos susceptibles, lo que resulta en una menor tasa de transmisión. Los individuos recuperados adquieren inmunidad, pudiendo perderla pasado cierto tiempo (típicamente tres años de acuerdo con estudios del SARS [3])

### b. Ecuaciones del modelo

El sistema de ecuaciones diferenciales ordinarias acopladas que gobierna la dinámica del virus del síndrome agudo respiratorio (SARS-COV-2) está dado por:

$$\frac{dS}{dt} = \Lambda - \frac{1}{N}S(\beta_{I1}I_1 + \beta_{I2}I_2 + \beta_H m_H H + \beta_V m_V V) - \mu S + fR \, , \, ). \tag{1}$$

$$\frac{dE}{dt} = \frac{\beta}{N}S(I_1 + I_2 + m_H H + m_V V) - (\alpha + \mu)E \, , \tag{2}$$

$$\frac{dI_1}{dt} = \alpha E - (\gamma + \mu)I_1 \, , \tag{3}$$

$$\frac{dI_2}{dt} = \gamma(1 - p_1)I_1 - (\lambda + \mu)I_2 \, , \tag{4}$$

$$\frac{dH}{dt} = \gamma p_1 I_1 - wH \, , \tag{5}$$

$$\frac{dR}{dt} = \lambda I_2 + w(1 - p_2)H - (\mu + f)R \, , \tag{6}$$

$$\frac{dV}{dt} = \varepsilon_1 I_1 + \varepsilon_2 I_2 + \varepsilon_3 H - \sigma V, \tag{7}$$

$$\frac{dM}{dt} = \mu(S + E + I_1 + I_2 + R) + wp_2 H \, , \tag{8}$$

donde cada grupo es mutuamente excluyente, y la suma de todos es:
$$N(t) = S(t) + E(t) + I_1(t) + I_2(t) + H(t) + R(t) \, . \tag{9}$$

El sistema de ecuaciones ha sido resuelto utilizando MATLAB Simulink con el método de solución numérica Bogacki-Shampine [12], un método Runge-Kutta de orden tres con cuatro etapas con la Primera mismo que la última (FSAL) Propiedad.

### c. Diagrama de transferencia

El diagrama del modelo (Figura 1) ilustra la progresión del coronavirus de un subgrupo a otro de la población del Paraguay.

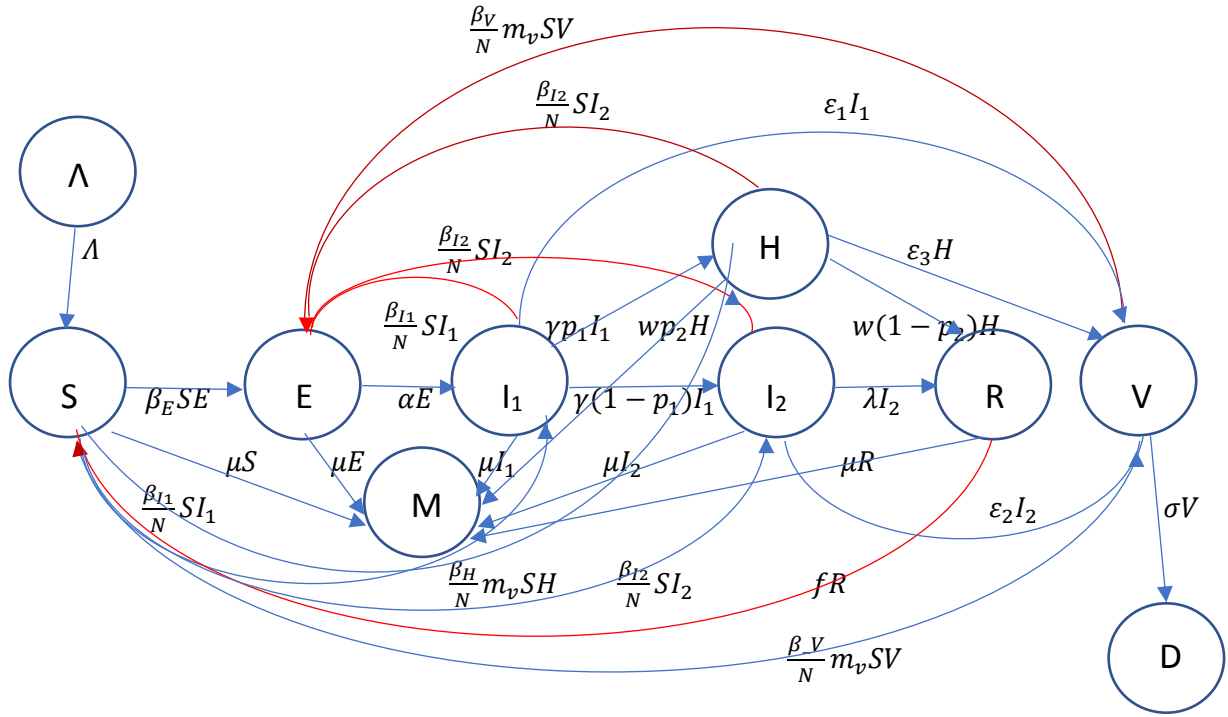

**Figura 1**. Diagrama de transferencia del SARS-CoV-2 en Paraguay

### d. Coeficientes de transmisión variables

Asumiendo que $\beta_{I1}(I_1)$, $\beta_{I2}(I_2)$ $\beta_H(H)$, $\beta_V(V)$ son positivos, y $\beta'_{I1}(I_1) \leq 0$, $\beta'_{I2}(I_2) \leq 0$ $\beta_H'(H) \leq 0$, $\beta_V'(V) \leq 0$, siendo funciones decrecientes, ya que valores mas altos para $I_1$, $I_2$, H y V motivan medidas de control mas fuertes, que reducen las tasas de transmisión.

$$\beta_{I1}(I_1) = \frac{\beta}{1+cI_1}, \tag{10}$$

$$\beta_{I2}(I_2) = \frac{\beta}{1+cI_2}, \tag{11}$$

$$\beta_H(H) = \frac{\beta}{1+cH}, \tag{12}$$

$$\beta_V(V) = \frac{\beta}{1+cV}. \tag{13}$$

### e. Cálculo del número básico de reproducción

El número básico de reproducción es definido como el número promedio de infecciones secundarias producidas cuando un individuo infectado es introducido en una población completamente susceptible **[13]**. Cinco compartimientos, E, $I_1$, $I_2$, H y V caracterizan el total de la población infectada por coronavirus en Paraguay.

Los términos de transmisión del sistema son $\frac{1}{N}S(\beta_{I1}I_1 + \beta_{I2}I_2 + \beta_H m_H H + \beta_V m_V V)$.

La matriz de nuevas infecciones F está dada por:

$$F = \begin{bmatrix} 0 & \frac{\beta_{I1}(0)}{N}S_0 & \frac{\beta_{I2}(0)}{N}S_0 & \frac{\beta_H(0)}{N}m_h S_0 & \frac{\beta_V(0)}{N}m_v S_0 \\ 0 & 0 & 0 & 0 & 0 \\ 0 & 0 & 0 & 0 & 0 \\ 0 & 0 & 0 & 0 & 0 \\ 0 & 0 & 0 & 0 & 0 \end{bmatrix}, \quad (14)$$

Los términos de transición del sistema son: $((\alpha + \mu)E)$, $(\alpha E - (\gamma + \mu)I_1)$, $(\gamma(1 - p_1)I_1 - (\lambda + \mu)I_2)$, $(\gamma p_1 I_1 - wH)$, $(\varepsilon_1 I_1 + \varepsilon_2 I_2 + \varepsilon_3 H - \sigma V)$.

La matriz de transición V está dada por:

$$V = \begin{bmatrix} \alpha + \mu & 0 & 0 & 0 & 0 \\ -\alpha & \gamma + \mu & 0 & 0 & 0 \\ 0 & -\gamma(1 - p_1) & \lambda + \mu & 0 & 0 \\ 0 & -\gamma p_1 & 0 & w & 0 \\ 0 & -\varepsilon_1 & -\varepsilon_2 & -\varepsilon_3 & \sigma \end{bmatrix}. \quad (15)$$

El número básico de reproducción es el radio espectral o el mayor valor propio dominante de la matriz de nueva generación $FV^{-1}$ [14]. Para el modelo descrito por las ecuaciones (1)-(13), obtenemos:

$$\mathfrak{R}_0 = \rho(FV^{-1}) = \frac{S_0}{N}\left[\beta_{I1}(0)\frac{\alpha}{(\alpha+\mu)(v+\mu)} + \beta_{I2}(0)\frac{\alpha\gamma(1-p_1)}{(\alpha+\mu)(v+\mu)(\lambda+\mu)} + \beta_H(0)m_H\frac{\alpha\gamma p_1}{w(\alpha+\mu)(v+\mu)} + \beta_V(0)m_v\frac{\varepsilon_1 \alpha w(\lambda+\mu)+\varepsilon_2 \alpha\gamma w(1-p_1)+\varepsilon_3 \alpha\gamma p_1(\lambda+\mu)}{(\alpha+\mu)(v+\mu)(\lambda+\mu)w\sigma}\right] = \mathfrak{R}_1 + \mathfrak{R}_2 + \mathfrak{R}_3 + \mathfrak{R}_4. \quad (16)$$

## III. Resultados
### a. Ajuste de curvas del modelo con datos reportados en Paraguay

Para estimar los valores de los parámetros $\beta$, $\varepsilon_1$, $m_v$ y c, ajustamos la curva del grupo infeccioso $I_2$ del modelo a los datos de casos positivos, hospitalizados y recuperados reportados diariamente por el Ministerio de Salud Pública y Bienestar Social de Paraguay [15] a partir del 6 de marzo hasta el 16 de abril de 2020. Para el ajuste de curva se ha utilizado el algoritmo Levenberg-Marquardt [16][17]. asignando las condiciones iniciales: $(S(0), E(0), I_1(0), I_2(0), H(0), R(0), V(0), M(0)) = (6.811x10^6, 1,1,1,0,0,0,0)$, basado en los datos reportados para el 6 de marzo. La probabilidad que los individuos infectados $I_2$ en cuarentena estricta y los hospitalizados H propaguen el coronavirus al medio ambiente conectado al público en general es muy baja, por lo cual $\varepsilon_2 = \varepsilon_3 = 0$ [4]. El tiempo infeccioso promedio es 15.20 días [2], siendo la suma de 1/γ y 1/λ.

A través del ajuste de datos, obtenemos una estimación del número de reproducción básico $R_0$= 4.68, muy cercano a valores reportados en la literatura, en el rango 2.7-8 [2]. Específicamente, $R_1$ = 1.50, $R_2$ = 2.44, $R_3$ = 0.07, $R_4$ = 0.67, lo que cuantifica el riesgo de infección a través de cada una de las cuatro vías. El mayor componente es $R_2$, la transmisión de los infecciosos no hospitalizados $I_2$ a los susceptibles, ya que presentan síntomas leves y pueden transmitir fácilmente a otras personas con las que entran en contacto. Le sigue en importancia $R_1$, correspondiente a la transmisión de los recientemente infecciosos $I_1$. El componente más pequeño $R_3$ proviene de la transmisión de los hospitalizados a los susceptibles, ya que se encuentran aislados de la sociedad. Observamos que $R_4$ muestra una contribución significativa del medio ambiente al riesgo de infección.

La figura 2 muestra el número de casos cumulativos confirmados positivos, recuperados y hospitalizados durante el periodo mencionado con las curvas $I_2$, R y H ajustadas del modelo.

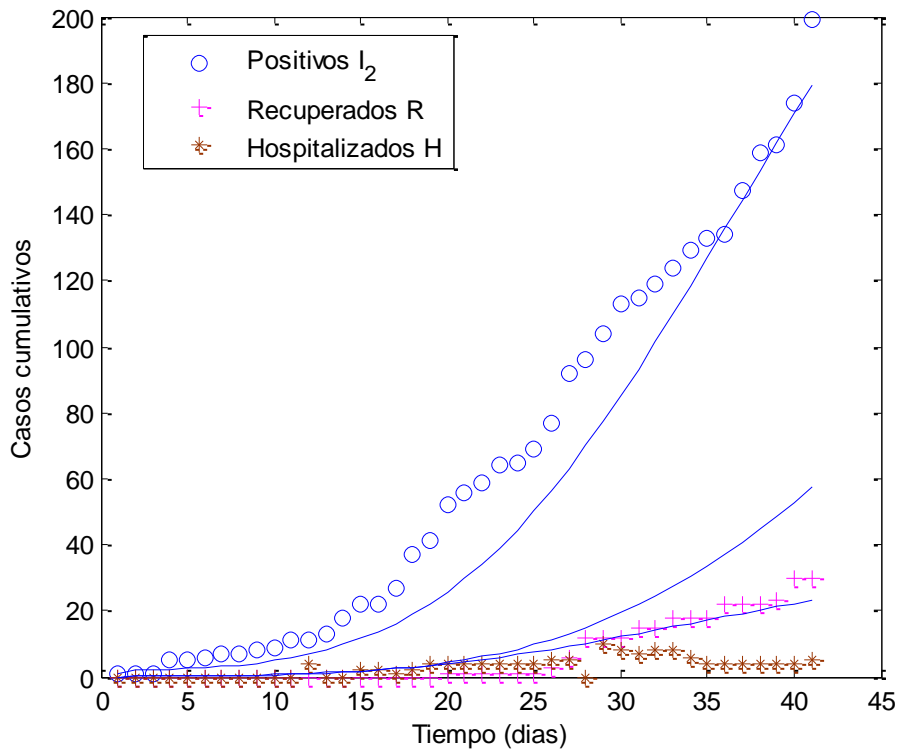

**Figura 2**. Casos cumulativos confirmados para Paraguay del 6 de marzo al 16 de abril de 2020. Los símbolos denotan los casos reportados positivos, recuperados, y hospitalizados; las líneas solidas denotan los resultados correspondientes de la simulación ($I_2$, R, H). El numero básico de reproducción es $R_0$=4.68. basado en los parámetros tabulados y el ajuste de curvas.

### b. Progresión de la pandemia sin adoptar medidas de control

Se han realizado pruebas numéricas usando tasas de transmisión simples y constantes en el modelo, es decir:

$$\beta_{I1}(I_1) = \beta_{I2}(I_2) = \beta_H(H) = \beta_V(V) = \beta, \tag{17}$$

equivalente a asignar c=0 en las ecuaciones (10)-(13). Se observa un nivel de infección muy elevado. En particular, se dan picos de infección entre los 80 y 120 días de pandemia, alcanzando E≈$1.9 \times 10^6$, $I_1$≈$1.2 \times 10^6$ e $I_2$≈$1.5 \times 10^6$ casos. El número de hospitalizados llega a un pico de 200.000 casos a los 100 días. Los resultados demuestran que las tasas de transmisión fijas, que no tienen en cuenta las medidas de control, pueden sobreestimar la severidad de la pandemia. También resaltan la importancia de adoptar medidas pertinentes ante la presencia del nuevo coronavirus.

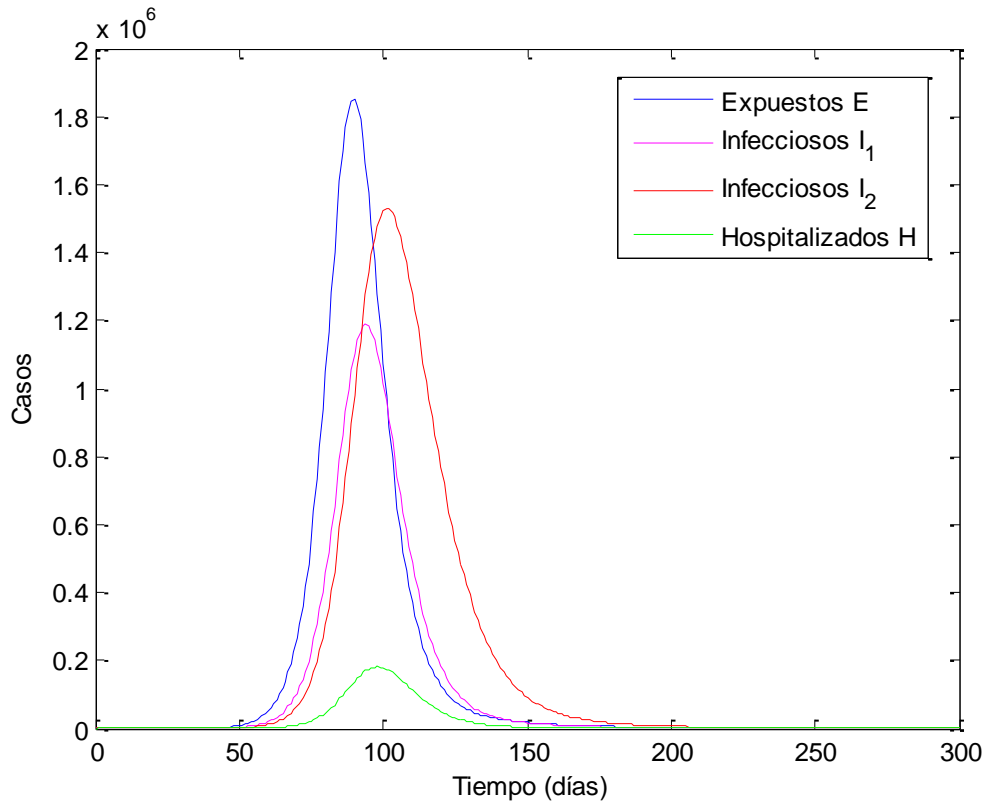

**Figura 3**. Resultados de la simulación del tamaño del brote de coronavirus en Paraguay usando las tasas de transmisión constantes (Ecuación 17). los parámetros tabulados y los resultados del ajuste de curvas.

### c. Progresión de la pandemia considerando las medidas adoptadas por el gobierno

Se ha realizado una predicción a corto plazo por 300 días usando el modelo con tasas de transmisión variables formuladas en las ecuaciones (10)-(13), que tienen en cuenta las medidas de control adoptadas por el gobierno, siendo más representativo de la actual situación de la pandemia en Paraguay. La figura 4 muestra que el nivel de infección, comenzando el 6 de marzo y marcado como día cero en la Figura 4, continuará creciendo por 150 días, llegará a un pico con alrededor de 233.000 infectados (E=80.000, I1=58.000 e I2 = 95.000, y luego descenderá gradualmente. El numero de hospitalizados llegara a un pico de 1.000 por día aproximadamente, incluyendo tanto los que presentan síntomas leves como los que están en terapia intensiva. Los avances actuales sobre el COVID-19 especulan que la enfermedad persistirá en el mundo humano y se volverá endémica, lo cual es demostrado por el modelo matemático y los resultados numéricos presentados, haciendo necesarias medidas preventivas y programas de intervención a largo plazo, hasta que estén disponibles nuevas vacunas y mejores tratamientos que están actualmente en desarrollo e investigación.

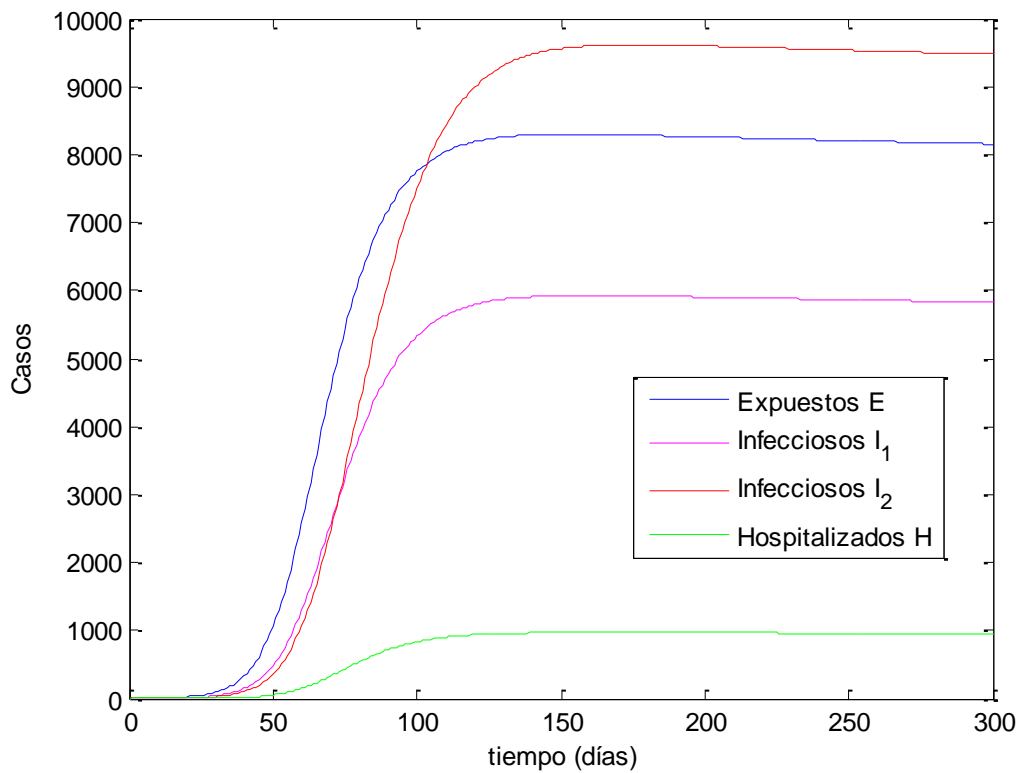

**Figura 4.** Resultados de la simulación del tamaño del brote de coronavirus en Paraguay usando las tasas de transmisión variables formuladas en las ecuaciones (10)-(13). los parámetros tabulados y los resultados del ajuste de curva.

## d. Análisis de sensibilidad de parámetros y limitaciones del modelo

El análisis de sensibilidad demuestra que las intervenciones, tales como el seguimiento de contactos intensivo seguido de cuarentena, aislamiento y restricciones de viajes, reducen efectivamente el riesgo de transmisión. En el modelo, el valor del coeficiente de ajuste de transmisión (c) en las ecuaciones (10)-(13) corresponde al grado de restricciones implementadas; a menor valor de c, aumenta la cantidad de infectados, volviéndose la curva más rápidamente exponencial. Con las medidas gubernamentales adoptadas, el número de infectados disminuye en 95%, comparado con el escenario sin restricciones. La tasa de transmisión básica ($\beta$) es proporcional al número de individuos infectados. La tasa de excreción del virus por personas inicialmente infecciosas ($\varepsilon_1$) y el factor de transmisión por virus en el medio ambiente ($m_v$) están directamente ligados a la cantidad de virus en el ambiente; valores mayores incrementan la cantidad de infectados y la propagación de la enfermedad

El modelo presentado posee las siguientes limitaciones: i) no está estructurado de acuerdo con la edad, ii) la presunción que el periodo de latencia es igual al periodo de incubación puede resultar en una sobreestimación del primero, y iii) la escasez de pruebas realizadas en Paraguay para detectar el Covid-19, las cuales son selectivas pudiendo producir un subregistro de casos

## IV. Conclusiones

Los resultados numéricos de las simulaciones demuestran la aplicación del modelo al brote de COVID-19 en Paraguay. El modelo se ajusta bien a los datos reportados por el ministerio de Salud Publica y Bienestar Social de Paraguay. A través del ajuste de datos, obtenemos una estimación del número de reproducción básico $R_0$= 4.68, el cual consiste en cuatro partes, representando las cuatro vías de transmisión: de los dos grupos de infecciosos, los individuos hospitalizados y el medio ambiente a los individuos susceptibles. Estos cuatro modos de transmisión conforman el riesgo de enfermedad de la pandemia, sugiriendo que las estrategias de intervención deben dirigirse a todas estas vías de transmisión. Las simulaciones consideran tasas de transmisión variables prediciendo la aparición de un pico pasado los 100 días del inicio de la pandemia, tras lo cual se reduce gradualmente. Usando tasas de transmisión constantes, los picos son mucho más elevados. Se especula que la enfermedad persistirá en el mundo humano y volverá endémica hasta que se desarrollen nuevas vacunas o mejores tratamientos., lo cual soporta el modelo matemático y resultados numéricos.

## Referencias